%
%
%
%
%
%
%
%
%
%
\hoffset=0.0in
\voffset=0.0in
\hsize=6.5in
\vsize=8.9in
\normalbaselineskip=12pt
\normalbaselines
\topskip=\baselineskip
\parindent=15pt
%
%
%

\let\ep=\epsilon

\let\lm=\lambda

\let\la=\langle
\let\ra=\rangle

\let\lf=\left
\let\rt=\right
\let\dt=\cdot
\let\del=\nabla

\let\q=\widehat

\let\h=\hbar

\let\rta=\rightarrow

\let\x=\times
\let\dy=\displaystyle
\let\ty=\textstyle
\let\sy=\scriptstyle
\let\ssy=\scriptscriptstyle
\let\:=\>
\let\\=\cr
\let\emph=\e

\let\m=\hbox

\let\cl=\centerline

\def\e#1{{\it #1\/}}
\def\textbf#1{{\bf #1}}
\def\[{$$}
\def\]{\[}
\def\re#1#2{$$\matrix{#1\cr}\eqno{({\rm #2})}$$}
\def\de#1{$$\matrix{#1\cr}$$}

\def\eqdf{\buildrel{\rm def}\over =}
\def\hf{{\sy {1 \over 2}}}

\def\hfs{{\ssy {1 \over 2}}}

\def\rg{r_{{\ssy >}}}

\def\qK{\q{K}}
\def\qV{\q{V}}
\def\qX{\q{X}}

\def\mathrm#1{{\rm #1}}

\def\mathcal#1{{\cal #1}}

\def\cF{{\cal F}}

\def\mbf{\fam\bffam\tenbf}
\def\bv#1{{\mbf #1}}

\def\vr{\bv{r}}

\def\vd{\bv{d}}
\def\vp{\bv{p}}

\def\vz{\bv{0}}

\def\qvp{\q{\vp}}

\def\Schr{Schr\"{o}\-ding\-er}
\font\frtbf = cmbx12 scaled \magstep1
\font\twlbf = cmbx12
\font\ninbf = cmbx9
\font\svtrm = cmr17
\font\twlrm = cmr12
\font\ninrm = cmr9
\font\ghtrm = cmr8

\def\gr#1{{\ghtrm #1}}

\def\abstract#1{{\ninbf\cl{Abstract}}\medskip
\openup -0.1\baselineskip
{\ninrm\leftskip=2pc\rightskip=2pc\noindent #1\par}
\normalbaselines}
\def\sct#1{\vskip 1.33\baselineskip\noindent{\twlbf #1}\medskip}

\def\so{\raise 0.65ex \m{\sevenrm 1}}
\def\sk{\par\vskip 0.66\baselineskip}
{\svtrm
\cl{A Self-Gravitational Upper Bound on Localized Energy,}
\medskip
\cl{Including that of Virtual Particles and Quantum Fields,}
\medskip
\cl{which Yields a Passable ``Dark Energy'' Density Estimate}
}
\bigskip
{\twlrm
\cl{Steven Kenneth Kauffmann\footnote{${}^\ast$}{\gr{Retired, American
Physical Society Senior Life Member, E-mail: SKKauffmann@gmail.com}}}
}
\bigskip\smallskip
\abstract{%
The self-gravitational correction to a localized spherically-symmetric
static energy distribution is obtained from energetically self-consis%
tent upgraded Newtonian gravitational theory.  The result is a gravi%
tational redshift factor that is everywhere finite and positive, which
both rules out gravitational horizons and implies that the self-gravi%
tationally corrected static energy contained in a sphere of radius $r$
is bounded by $r$ times the fourth power of $c$ divided by $G$.  Even
in the absence of spherical symmetry this energy bound still applies
to within a factor of two, and it cuts off the mass deviation of any
quantum virtual particle at about a Planck mass.  Because quantum un%
certainty makes the minimum possible energy of a quantum field infin%
ite, such a field's self-gravitationally corrected energy attains the
value of that field's containing radius $r$ times the fourth power of
$c$ divided by $G$.  Roughly estimating any quantum field's containing
radius $r$ as $c$ times the age of the universe yields a ``dark ener%
gy'' density of 1.7 joules per cubic kilometer.  But if $r$ is put to
the Planck length appropriate to the birth of the universe, that ener%
gy density becomes the enormous Planck unit value, which could con%
ceivably drive primordial ``inflation''.  The density of ``dark ener%
gy'' decreases as the universe expands, but more slowly than the den%
sity of ordinary matter decreases.  Such evolution suggests that
``dark energy'' has inhomogeneities, which may be ``dark matter''.
}

\sct{Introduction: self-gravitational capping of net localized energy}
\noindent
Einstein's identification of mass as a form of energy and of energy as
a source of gravitation has the basic, but regrettably little appreci%
ated, consequence of capping the possible energy content of any local%
ized region at a value which is proportional to $(c^4/G)$ times that
region's maximum linear dimension.  A very rough heuristic picture of
how this comes about can be gleaned from thinking about two equal
idealized ``point masses'' of value $m > 0$ that are separated by a
distance $d > 0$.  Of course these two ``point masses'' will gravita%
tionally attract each other with a vector force $-(Gm^2\vd)/|\vd|^3$,
where $|\vd| = d$, which tends to reduce their separation.  We can
either impose a constraint to stop that from occurring, or, a bit more
elegantly, kick the two particles into a gravitational mutual circular
orbit which automatically maintains their desired relative separation
$d$. The kinetic energy of such a circular orbit is half of the nega%
tive of its potential energy $-(Gm^2)/d$ from the virial theorem~[1],
so that with the inclusion of the energy of the two masses the sys%
tem's total energy comes to $[2mc^2 - ((Gm^2)/(2d))]$.  As a function
of the value $m$ of those two equal ``point masses'', this total ener%
gy \e{has a maximum value} which occurs at $m = 2(c^2/G)d$ and equals
$2(c^4/G)d$.  In view of the fact \e{that we haven't made an allow%
ance} for the negative gravitational energy \e{which arises as a con%
sequence of the system's orbital kinetic energy}, this result of
$2(c^4/G)d$ \e{is actually an overestimate} of the system's maximum
possible net effective energy.

If we \e{alternatively} simply \e{impose} the separation $d$ of the
two particles \e{as a static constraint}, the system's total energy
\e{instead} comes to,
\re{
E(m; d) = [2mc^2 - ((Gm^2)/d)],
}{1a}
which as a function of $m$ has a maximum value that occurs at $m =
(c^2/G)d$ and equals,
\re{
E_{{\rm max}}(d) = (c^4/G)d.
}{1b}

It is therefore apparent \e{that in any case} the self-gravitation of
such a \e{localized} system indeed \e{caps its net effective energy}
at a value which is proportional to $(c^4/G)$ times its maximum linear
dimension $d$.  Now such an energy cap proportional to a system's max%
imum linear dimension implies that any ``point'' system must have
\e{zero} energy!  So our use of two equal \e{nonzero} ``point masses''
to \e{construct} the systems of the above examples \e{isn't really
physically self-consistent}.  The next section will therefore deal
with \e{nonsingularly distributed localized positive energy}, albeit
only for such localized positive energy distributions which are static
and spherically symmetric; more general nonsingularly distributed
static localized positive energy is taken up in a later section.  The
fact that \e{nonzero point masses do not exist in relativistic gravity
theory} has been very strongly emphasized \e{as well} by Christoph
Schiller in connection with his demonstration of the equivalence of
his principle of maximum force to the Einstein equation in a context
of general covariance~[2].

\sct{Self-gravitational reduction of spherically-symmetric localized
static energy}
\noindent
We denote an initially prescribed \e{idealized} (i.e., \e{before} its
self-gravitational energy reduction) static spher\-i\-cal\-ly-symmet%
ric nonsingular positive-energy density that is localized to a sphere
of radius $r_s$ as $T_{G = 0}(r)$, where, $T_{G = 0}(r) > 0$ for $r <
r_s$ and $T_{G = 0}(r) = 0$ for $r > r_s$, while we denote the corres%
ponding \e{physically realistic} self-gravitationally reduced \e{ef%
fective} energy density as $T_G(r)$.  These two energy densities
\e{are related to each other through the dimensionless local negative
Newtonian gravitational potential} $\phi(r)$ \e{via} $T_G(r) = T_{G =
0}(r)(1 + \phi(r))$.  Furthermore, Newtonian gravitational theory
\e{is upgraded to being energetically self-consistent} when $\phi(r)$
follows from the physically realistic self-gravitationally reduced ac%
tually \e{effective} energy density $T_G(r)$ via the Newtonian expres%
sion,
\re{
{\ty \phi(r) = \phi(|\bv r|) =
{\sy -(G/c^4)}\int_{|\bv r'|\leq r_s}
{d^3\bv r' T_G(|\bv r'|)\over|\bv r - \bv r'|} =
{\sy -((4\pi G)/c^4)}\int_0^{r_s}{\sy (r')^2dr'T_G(r')/\rg},}
}{2a}
where $\rg$ abbreviates $\max(r, r')$.  It is actually \e{more conven%
ient} to deal \e{directly} with the dimensionless \e{local gravita%
tional energy-reduction factor} $\psi(r)\eqdf(1 +\phi(r))$ than with
the local Newtonian gravitational \e{potential} $\phi(r)$.  Transcrib%
ed in terms of $\psi(r)$ \e{and the initially prescribed idealized en%
ergy density} $T_{G = 0}(r)$, Eq.~(2a) becomes,
\re{
{\ty\psi(r) = 1 - ((4\pi G)/c^4)\int_0^{r_s}(r')^2dr'T_{G = 0}(r')
\psi(r')/\rg,}
}{2b}
which transparently manifests the \e{fed back} character of $\psi(r)$;
\e{elementary} Newtonian gravitational theory which is \e{not} ener%
getically self-consistent \e{is recovered} when $\psi(r')$ within the
integrand on the right-hand side of Eq.~(2b) \e{is replaced by unity}.
We further note that since,
\re{
T_G(r) = T_{G = 0}(r)\psi(r),
}{2c}
Eq.~(2b) is readily verified to imply that,
\re{
\hbox{$\psi(r) = 1 -(r_s/r)((GE_G)/(c^4 r_s))$ when $r\geq r_s$,}
}{2d}
where $E_G$ is of course the spherically-symmetric static energy dis%
tribution's self-gravitationally reduced \e{effective total energy}.

For a localized photon wave packet the dimensionless local gravita%
tional energy-reduction factor $\psi(r)$ is obviously \e{also} a local
gravitational \e{frequency reduction factor}.  Therefore its dimen%
sionless \e{inverse} $(1/\psi(r))$ \e{is the local gravitational red%
shift or time-dilation factor}.  We thus are here \e{implicitly} deal%
ing with the $g_{00}(r)$ component of a metric tensor~[3],
\re{
g_{00}(r) = (\psi(r))^2.
}{2e}
Since $\psi(r)$ is based on the energetically self-consistent upgraded
\e{Newtonian} gravitational potential $\phi(r)$ of Eq.~(2a), our
\e{implicit} metric tensor that is alluded to in Eq.~(2e) is expressed
in ``Newtonian'' coordinates.

In order to understand the \e{behavior} of $\psi(r)$ we now proceed to
\e{repeatedly differentiate} Eq.~(2b).  To that end we explicitly ex%
press $(1/\rg)$ as $\theta(r - r')/r + (1 -\theta(r - r'))/r'$, and
thus obtain $d(1/\rg)/dr = -\theta(r - r')/r^2$.  Applying this result
to Eq.~(2b) yields,
\re{
{\ty d\psi(r)/dr = ((4\pi G)/(c^4r^2))\int_0^{\min(r, r_s)}(r')^2dr'
T_{G = 0}(r')\psi(r'),}
}{2f}
Furthermore, $d^2(1/\rg)/dr^2 = -\delta(r - r')/r^2 - (2/r)d(1/\rg)/
dr$, which applied to Eq.~(2b) yields,
\re{
d^2\psi(r)/dr^2 + (2/r)d\psi(r)/dr = \theta(r_s - r)((4\pi G)/c^4)
T_{G = 0}(r)\psi(r).
}{2g}
Eq.~(2g) reveals that $\psi(r)$ satisfies a zero-energy S-wave
Schr\"{o}\-din\-ger-equa\-tion analogue with a potential barrier pro%
duced by $T_{G = 0}(r)\geq 0$ in the region $0\leq r\leq r_s$ and emp%
ty space in the region $r > r_s$, where the solution $\psi(r)$ is ana%
lytically represented by Eq.~(2d).  Given the fact of zero energy, the
potential barrier resists penetration anywhere that it is positive, so
for $0\leq r\leq r_s$ we would expect $\psi(r)$ \e{to approximately
increase exponentially with increasing} $r$.

This behavior is \e{indeed} manifested by the ``within the barrier''
S-wave WKB approximation,
\de{
\hbox{$\psi(r)\approx (r_0/r)\sinh\left(\int_0^r dr'[((4\pi G)/c^4)
T_{G = 0}(r')]^\hfs\right)$ for $0\leq r\leq r_s$,}
}
which is to be smoothly joined to Eq.~(2d) at $r = r_s$, \e{which
would approximately determine both the positive constant} $r_0$ \e{and
also the self-gravitationally reduced effective total energy} $E_G$
that is needed to make Eq.~(2d) definite.  By thus utilizing the
``within the barrier'' WKB approximation together with Eq.~(2d) we ob%
tain a cobbled-together picture of the behavior of $\psi(r)$, namely
\e{an increasing positive function of} $r$ \e{for all} $r\geq 0$
\e{which smoothly rises from its minimum positive value at} $r = 0$
\e{toward its asymptotic value of unity as} $r\rta\infty$.

By enlisting the help of Eq.~(2f) we can \e{reaffirm} this picture of
$\psi(r)$ \e{without utilizing the WKB approximation}.  Eq.~(2f) re%
lates the local $r\rta 0+$ behaviour of $d\psi(r)/dr$ to $\psi(r = 0)$
as follows,
\re{
d\psi(r)/dr \sim ((4\pi G)/(3c^4))r T_{G = 0}(r = 0)\psi(r = 0),
}{3a}
which implies that as $r\rta 0+$,
\re{
\psi(r) \sim [1 + ((2\pi G)/(3c^4))r^2 T_{G = 0}(r = 0)]\psi(r = 0).
}{3b}
Assuming that $T_{G = 0}(r = 0) > 0$, we see from Eq.~(3b) that if
$\psi(r = 0) < 0$, then $\psi(r)$ will initially \e{decrease from that
negative value}, which makes the Eq.~(2f) expression for $d\psi(r)/dr$
\e{negative}, the upshot being a negative $\psi(r)$ \e{that keeps on
decreasing}.  If $\psi(r = 0) = 0$ we see from Eq.~(3b) that \e{the
first and second derivatives of} $\psi(r)$ \e{at} $r = 0$ \e{also van%
ish}.  Therefore, given the linear homogeneous second-order differen%
tial Eq.~(2g), $\psi(r)$ \e{vanishes identically all the way to} $r =
r_s$, which behaviour \e{can't be smoothly joined there to the}
$\psi(r)$ \e{form given by} Eq.~(2d).  Therefore, the $r\rta\infty$
\e{asymptotic value of} $+1$ \e{for} $\psi(r)$ \e{that} Eq.~(2d)
\e{requires} is compatible \e{only} with a \e{positive} value for
$\psi(r = 0)$, from which Eq.~(2f) tells us $\psi(r)$ \e{keeps on
increasing}.  In cases that $T_{G = 0}(r = 0)$ \e{vanishes}, very
\e{similar} arguments regarding the positive and increasing nature of 
$\psi(r)$ apply for all values of $r$ larger than that $r$-value
beyond which $T_{G = 0}(r)$ \e{first takes on positive values}, with
$\psi(r)$ \e{being a positive constant} for $r$-values \e{smaller}
than that.

Having thereby shown that,
\re{
\hbox{$1\geq\psi(r) > 0$ for all $r\geq 0$,}
}{4a}
we note that this inequality holds in particular at $r = r_s$, which
from Eq.~(2d) implies that,
\re{
0\leq ((GE_G)/(c^4 r_s)) < 1.
}{4b}

Finally, if one \e{increases} $T_{G = 0}(r)$, it is apparent from 
Eq.~(2b) that the positive increasing function $\psi(r)$ responds
by \e{decreasing} in the interval $0\leq r\leq r_s$.  From Eq.~(2b)
one sees that the average value of $\psi(r)$ in the interval
$0\leq r\leq r_s$ times roughly $[1 + ((GE_{G = 0})/(c^4 r_s))]$
is equal to unity.  Therefore \e{multiplying} $E_{G = 0}$ by the
dimensionless factor $N\geq 1$ roughly \e{divides} the average
value of $\psi(r)$ in the interval $0\leq r\leq r_s$ by the factor
$[1 + N((GE_{G = 0})/(c^4r_s))]/[1 + ((GE_{G = 0})/(c^4r_s))]$.

In any event, taking $E_{G = 0}\rta\infty$ produces \e{an impenetrable
barrier} in the interval $0\leq r\leq r_s$, which implies that,
\re{
\hbox{$\psi(r)\rta 0$ for $0\leq r\leq r_s$ when $E_{G = 0}\rta\infty
$.} 
}{4c}
For $r = r_s$, Eq.~(4c) together with Eq.~(2d) imply that,
\re{
\hbox{$((GE_G)/(c^4 r_s))\rta 1$ when $E_{G = 0}\rta\infty$.}
}{4d}

The inequality $\psi(r) > 0$ of Eq.~(4a) \e{rules out gravitational
horizons}---which occur where the gravitational redshift factor
$(1/\psi(r))$ is locally infinite---\e{regardless of the strength of
the nonnegative} $T_{G = 0}(r)$.  However, Eq.~(4c) tells us that the
positive \e{finite} gravitational local redshift factor $(1/\psi(r))$
can certainly be made \e{arbitrarily large}.  These \e{dual aspects}
of the behavior of the gravitational local redshift factor are related
to the fact that Eq.~(2b) \e{feeds back that factor's inverse}
$\psi(r)$.

The issue of whether gravitational horizons can \e{ever} be physically
realized in General Relativity \e{was fundamentally settled in the
negative} by Christoph Schiller \e{in the course of his demonstration}
that his principle of the \e{unattainable} least upper bound of $(c^4
/(4G))$ on force magnitudes is, against a backdrop of general covar%
iance, \e{equivalent} to the Einstein equation~[2].

With regard to Schiller's principle vis-\`{a}-vis the inequalities
given by our Eqs.~(4a) and (4b), we note that the \e{Newtonian gravi%
tational force magnitude} between \e{two identical} localized spheri%
cally-symmetric static systems, each with radius $r_s$ and self-gravi%
tationally reduced effective total energy $E_G$ is equal to, \e{when
they just touch}, $G(E_G)^2/[c^4(2 r_s)^2]$.  Requiring this to be
\e{less} than Schiller's unattainable least upper bound $(c^4/(4G))$
\e{also} yields our key Eq.~(4b) inequality $((GE_G)/(c^4 r_s)) < 1$
\e{that reflects the nonexistence of a gravitational horizon at} $r =
r_s$.

Although Schiller's principle \e{rules out physical realization of
gravitational horizons}, the fact that physical systems can come
\e{arbitrarily close} to attaining horizons (which is explicitly noted
above in the paragraph below Eqs.~(4c) and (4d)) \e{actually plays a
crucial role in the demonstration that Schiller presents}~[2].

The key inequality $((GE_G)/(c^4 r_s)) < 1$ of Eq.~(4b) \e{also}
implies that if a localized static, spherically symmetric energy
distribution is shrunk to a \e{point}, i.e., if its radius $r_s$ is
taken to zero, then \e{that sphere's effective energy} $E_G$ \e{is
as well forced to zero}.  Therefore nonzero effective static \e{point
energies} indeed \e{don't exist}, as noted in the introductory section
of this paper---Schiller stresses that appreciation of that fact is
crucial to proper understanding of General Relativity~[2].

From Eq.~(2d) we see that in the \e{empty-space} Schwarzschild
region $r > r_s$ our ``Newtonian'' $g_{00}(r)$ of Eq.~(2e) has
the value $(1 - ((GE_G)/(c^4 r)))^2$.  Since the well-known
``isotropic'' Schwarzschild metric tensor~[4] has $g_{00}(\rho) =
(1 - ((GE_G)/(2c^4\rho)))^2/(1 + ((GE_G)/(2c^4\rho)))^2$, it can be
readily verified that the ``iso\-tro\-pic'' Schwarzschild metric
tensor is mapped into our ``Newtonian'' Schwarzschild metric tensor by
the simple coordinate transformation $\rho(r) = r - (GE_G)/(2c^4)$.
Applying that transformation to the \e{full} ``isotropic'' Schwarz%
schild metric tensor explicitly yields our \e{full} ``Newtonian''
Schwarzschild metric tensor,
\re{
ds^2  =  (1 - ((GE_G)/(c^4r)))^2(cdt)^2 -
           (1 - ((GE_G)/(2c^4r)))^{-4}dr^2 - \cr
         (1 - ((GE_G)/(2c^4r)))^{-2}((rd\theta)^2 +
                    (r\sin\theta d\phi)^2),
}{5}
\e{which has the  empty-space region of validity} $r > r_s$.  Since
$((GE_G)/(c^4 r_s)) < 1$ from Eq.~(4b), it follows that in this
``Newtonian'' Schwarzschild metric tensor's empty-space \e{region of
validity} $r > r_s$, $((GE_G)/(c^4r)) < 1$.  That inequality makes it
manifest that within its region of validity the Eq.~(5) ``Newtonian''
Schwarzschild metric tensor \e{has no horizon nor any other unphysical
anomaly}.

Of course \e{the reason that no horizon occurs} is that the self-%
gravitational \e{capping} of localized energy which is \e{inherent}
to Eq.~(4b) \e{always} locates the \e{putative} horizon of this
``Newtonian'' Schwarzschild metric tensor in a region that \e{is not
in fact in empty space}; i.e., that is \e{outside} of the \e{empty-%
space region of validity} of the ``Newtonian'' Schwarzschild metric
tensor.

That the putative horizon \e{always} ``occurs'' in a non-empty-space
region where the ``Newtonian'' Schwarz\-schild metric tensor \e{is
inherently invalid} is \e{indeed fortunate}, because \e{at its} $r =
(GE_G)/c^4$ \e{putative horizon} the Eq.~(5) ``Newtonian'' Schwarz%
schild metric tensor has bizarre (3 + 0) dimensions, \e{which violates
the metric-tensor signature theorem}~[5]!

\sct{Interacting quantum particles that have forbidden energy or mass}
\noindent
Because of its wave character, a quantum particle which interacts with
a potential can penetrate a short distance into a region where the po%
tential's value exceeds the particle's energy---such penetration is
forbidden to classical particles.  In such a region the quantum par%
ticle's kinetic energy and momentum squared effectively assume \e{neg%
ative} values, and its penetration length $\lm$ into that energetical%
ly forbidden region can be roughly described as,
\re{
    \lm \approx \h/(-p_{{\rm eff}}^2)^\hfs.
}{6a}
Since for a nonrelativistic interacting particle,
\re{
    E = p^2/(2m) + V,
}{6b}
we can rewrite Eq.~(6a) as,
\re{
    \lm\approx \h/(2m(V - E))^\hfs,
}{6c}
and the energetically forbidden region corresponds to $V > E$.  If $V$
should vary significantly from the edge of the forbidden region to the
depth $\lm$, Eq.~(6c) will need to be regarded as an approximate
\e{implicit relationship} which actually needs to be \e{solved} for
$\lm$.  That implicit relationship is set up by reexpressing $V$ \e{as
a function of the distance from the edge of the forbidden region}, and
the resulting new independent variable is then identified as $\lm$.

A highly energetic interacting quantum particle can similarly deviate
from the natural rest mass which it has when it is free, and thus en%
ter a region of forbidden rest mass.  Just as there is an effective
length limit for quantum particle penetration into a region of forbid%
den energy, so there is an effective proper time limit, i.e., life%
time, for quantum particle penetration into a region of forbidden rest
mass, namely,
\re{
    \tau \approx \h/(\Delta_m c^2),
}{7a}
where $\Delta_m$ is the rest-mass deviation experienced by an energet%
ic interacting quantum particle and $\tau$ is the lifetime of that de%
viant-mass state.  Given its limited lifetime, such a deviant-mass
\e{virtual particle} is as well \e{limited in space} to a spherical
region of radius $R\approx c\tau$,
\re{
    R\approx c\tau\approx \h/(\Delta_m c).
}{7b}
Given that radius $R$ of Eq.~(7b), Eq.~(4b) then yields a self-gravi%
tational approximate upper bound on the total energy $E$ of this dev%
iant-mass particle,
\re{
    E \>\widetilde{<}\> (c^4/G)R \approx c^3\h/(G\Delta_m).
}{7c}
Since $\Delta_m c^2$ cannot exceed the deviant-mass particle's total
energy $E$, it follows from Eq.~(7c) that,
\re{
    \Delta_m c^2 \>\widetilde{<}\> c^3\h/(G\Delta_m).
}{7d}
Therefore,
\re{
    (\Delta_m)^2 \>\widetilde{<}\> \h c/G,
}{7e}
which implies that,
\re{
    \Delta_m \>\widetilde{<}\> (\h c/G)^\hfs,
}{7f}
namely that $\Delta_m$ is approximately bounded by the \e{Planck mass}
$(\h c/G)^\hfs$.

This approximate Planck-mass upper bound on the mass deviation of any
interacting quantum virtual particle universally cuts off the \e{ul%
traviolet divergences} which bedevil quantum particle scattering am%
plitude calculations.

\sct{Self-gravitational reduction of the infinite energies of quantum
fields}
\noindent
A noninteracting field (e.g., the source-free electromagnetic field)
always decomposes into an infinite number of independent simple har%
monic oscillators whose frequency spectrum has no upper bound.  Upon
quantization, each such oscillator has a \e{minimum positive energy}
(i.e., quantum ground state energy) which is equal to its frequency
times $\h/2$: that minimum energy is quantum theoretically \e{inviola%
ble}, being \e{completely mandated by the quantum uncertainty princi%
ple}.  The fact that there are an \e{infinite number} of such simple
harmonic oscillators with an \e{unbounded frequency spectrum} implies
that the corresponding quantum field \e{always has infinite energy}.
If the field is set up in a bounded region, then its quantum counter%
part has \e{not only} infinite energy, but necessarily \e{infinite
average energy density} as well.

Therefore for quantum fields the uncertainty principle baldly con%
fronts us with an unphysical nightmare, yet \e{without} this \e{self%
same} uncertainty principle \e{likewise} mandating a definite minimum
energy, Rutherford's nuclear atom can't be sustained.

The last apparent hope for the beleaguered theorist in this harrowing
circumstance lies with \e{the self-gravitational reduction of an in%
itially infinite energy}, $E_{G = 0}\rta\infty$, which is contained in
a spherical region of radius $r_s$, namely the situation described by
Eq.~(4d).  A grace note is that the self-gravitationally reduced ener%
gy \e{result} for that situation \e{is both finite and simple},
\re{
    E_G = (c^4/G)r_s.
}{8a}
Of course it makes no difference whatsoever \e{how many} or \e{what
types} of quantum fields are contained in that spherical region of ra%
dius $r_s$; \e{any infinite} initial energy contained in that region
produces the result of Eq.~(8a).

The \e{total} self-gravitationally reduced energy of quantum fields,
which is what Eq.~(8a) is supposed to describe, isn't \e{directly}
measurable.  However, by making use of the fields' containing radius
$r_s$, we can obtain from Eq.~(8a) their self-gravitationally reduced
\e{averaged energy density} $\bar\rho$,
\re{
    \bar\rho = (3c^4/(4\pi G r_s^2)).
}{8b}
It now remains to puzzle out what conceivable physics could produce
the quantum fields' containing radius $r_s$.  It is, of course, appar%
ent that \e{no material substance} can serve to contain \e{the arbi%
trarily high frequencies} which such fields are able to muster. The
universe' \e{cosmological redshift}, however, in principle ought to
defang \e{any} frequency, and indeed appears to serve as the ``con%
tainment'' for what we can possibly hope to survey.  Therefore a not
altogether implausible crude estimate of the quantum fields' ``con%
taining radius'' $r_s$ ought to be given by the age of the universe~%
[6] times the speed of light, which comes to about $1.3\x 10^{26}$ me%
ters.  Putting that value of $r_s$ into Eq.~(8b) yields about $1.7\x
10^{-9}$ joules per cubic meter (i.e., $1.7\x 10^{-8}$ ergs per cubic
centimeter or $1.7$ joules per cubic kilometer) as a crude estimate of
the universe' average ``dark energy'' density.  This is in fact of the
same order of magnitude as what is yielded by observations~[7, 8].

In addition to the ability of Eq.~(8b) to yield a passable crude esti%
mate of the current universe' average ``dark energy'' density, its
systematics also seem fascinating.  If we project it all the way back
to the universe' birth, when $r_s$ was presumbably of the order of
magnitude of the Planck length $(G\h/c^3)^\hfs$, then $\bar\rho$ ap%
proaches of order unity in Planck units of energy density, which is
roughly 120 orders of magnitude greater than its value for the current
universe.
 
Theorists \e{who did not attempt to actually model the physics which
produces self-gravitational energy correction} have favored this par%
ticular \e{enormous} value of ``dark energy'' density because of their
adoption of a physically-blinkered ``universal  fix'' for infinite
results, namely the replacement of any such infinity by one Planck
unit of the appropriate dimensions~[8].  Neither physical modeling of
self-gravitational energy correction nor observations have much over%
lap with such undiscriminating replacement of the quantum energy-den%
sity infinity by its Planck-unit value, but it is still fascinating
to consider that enormous  Planck unit of ``dark energy'' density as
being relevant to the \e{early} universe, as that would apparently
provide an \e{automatic} mechanism for the heretofore puzzling ``in%
flation'' of that early universe.

Finally, Eq.~(8b) suggests that the average ``dark energy'' density
ought to decrease toward zero as the universe continues its expansion.
This brings to mind the not infrequently expressed theorist preference
for \e{exactly vanishing} ``dark energy'' density over its \e{ob%
served} value~[8], which while \e{immensely smaller} than the Planck
unit of energy density, nonetheless \e{absolutely fails to vanish}.
In fact, completely to the \e{contrary}, it \e{dominates} the \e{net}
average energy density of our universe~[7].  We see that Eq.~(8b) ap%
parently caters for \e{all} tastes in average ``dark energy'' density,
whether those tastes gravitate toward the enormous  Planck unit of en%
ergy density, zero energy density, or anything in between, including a
passable rendition of the observed average ``dark energy'' density
which actually obtains at the current stage of evolution of our uni%
verse.  It is to be cautioned, however, that while Eq.~(8b) indeed has
average ``dark energy'' density decreasing toward zero as the universe
continues to expand, the average density of \e{normal} matter would be
expected to decrease \e{at a faster rate}, so that ``dark energy''
relative \e{dominance} would continue to grow.

\sct{Must self-gravitation be quantized to correct quantum energy in%
finities?}
\noindent
Gravity, like electromagnetism, is a \e{gauge theory}, and the issues
surrounding its quantization formally parallel those issues in elec%
tromagnetism.  In both cases there are dynamical, nondynamical and
redundant fields present, and the \e{dynamical} fields in \e{both}
cases are \e{two} in number and describe \e{transverse radiation}.
\e{Only these two dynamical radiation fields are subject to quantiza%
tion.}

What remains after the two transverse dynamical radiation fields are
accounted for splits evenly into nondynamical and redundant fields,
\e{neither} of which, of course, are subject to quantization.  The
four-potential of electromagnetism yields one redundant field and one
nondynamical field of Coulombic character.  The symmetric metric ten%
sor of gravity yeilds four redundant fields and four nondynamical
ones.  One of the nondynamical fields very roughly corresponds to New%
tonian gravity with roughly an energy-density source, while the other
three merely round out a relativistic four-vector representation, and
therefore have roughly a momentum-flux source.

If we look back at the previous parts of this article, it is clear
that the self-gravitational corrections which are of overarching im%
portance can all be profitably pondered in a quasi-static or outright
static framework.  The basic ingredients for self-gravitational cor%
rections tend to be \e{Newtonian}, albeit \e{an energetically self-%
consistent form} of gravitational Newtonianism.

Gravitational radiation \e{doesn't} physically enter into self-gravi%
tational correction, so gravity quantization \e{cannot} be an issue
in such correction, any more than electromagnetic quantization can be
an issue in electrostatics.  Gravitostatics is merely more subtle than
electrostatics because of its energetic self-consistency.

\sct{Self-gravitational reduction of arbitrary smooth localized static
energy}
\noindent
In the second section of this article we showed that any spherically-%
symmetric static nonsingular positive energy distribution of radius
$r_s$ has, after self-gravitational reduction, a cap on its total ef%
fective energy $E_G$ that is given by Eq.~(4b), namely $E_G <
(c^4/G)r_s$.

We now \e{extend} our study of the self-gravitational energy-reduction
process begun in the second section to \e{any} specified \e{static
energy density} $T_{G = 0}(\vr)$ which is nonnegative, smooth and
globally integrable, i.e.,
\re{
    T_{G = 0}(\vr)\geq 0,
}{9a}
\re{
    \hbox{$\del_{\vr}\lf(T_{G = 0}(\vr)\rt)$ is continuous,}
}{9b}
and,
\re{
    \int T_{G = 0}(\vr')d^3\vr' < \infty .
}{9c}
Here we \e{specifically refrain} from making the assumption of the
second section of this article that $T_{G = 0}(\vr)$ \e{possesses
spherical symmetry}, nor do we assume that it possesses \e{any other
particular symmetry}.

Now if it were the case that we \e{actually had in hand the self-grav%
itational correction} $T_G(\vr)$ of the specified static energy dens%
ity $T_{G = 0}(\vr)$, we could calculate the negative Newtonian gravi%
tational work done to bring the \e{infinitesimal original static ener%
gy} $T_{G = 0}(\vr)d^3\vr$ from infinity to its position at $\vr$
while subject to the static gravitational field that is provided by
$T_G(\vr)$, which yields the result $-(G/c^4)\int d^3\vr'T_G(\vr')|\vr
- \vr'|^{-1}T_{G = 0}(\vr)d^3\vr$.  However, because the static gravi%
tational interaction inherently occurs between \e{pairs} of infinites%
imal energies, we must take care to \e{avoid double-counting}, so we
assign only \e{half} of this \e{negative gravitational work correc%
tion} to the infinitesimal static energy located at $\vr$, and thus
obtain,
\re{
    T_G(\vr)d^3\vr = \lf[1 - \hf(G/c^4)\int d^3\vr'T_G(\vr')
    |\vr - \vr'|^{-1}\rt]T_{G = 0}(\vr)d^3\vr.
}{10a}
From Eq.~(10a) we see that the dimensionless \e{local gravitational
energy-reduction factor} $\cF_G(\vr)$ that satisfies,
\re{
    T_G(\vr)d^3\vr = \cF_G(\vr)T_{G = 0}(\vr)d^3\vr,
}{10b}
is given by,
\re{
    \cF_G(\vr)\eqdf 1 - \hf(G/c^4)\int|\vr - \vr'|^{-1}T_G(\vr')d^3\vr'.
}{10c}
This static local gravitational energy-reduction factor $\cF_G(\vr)$
is obviously the \e{inverse} of the corresponding gravitational time-%
dilation factor, and thus is equal to $(g_{00}(\vr))^\hfs$~[3].  If we
insert the instruction implicit in Eq.~(10b) into the right-hand side
of Eq.~(10c), we obtain the following inhomogeneous linear integral
equation for $\cF_G(\vr)$,
\re{
    \cF_G(\vr) = 1 - \hf(G/c^4)\int|\vr - \vr'|^{-1}T_{G = 0}(\vr')
     \cF_G(\vr')d^3\vr'.
}{10d}
In light of Eq.~(9c), we can deduce from Eq.~(10d) that,
\re{
    {\dy \lim_{|\vr|\rta\infty}}\cF_G(\vr) = 1.
}{11a}
Furthermore, since the integral transform \e{kernel} $-1/\lf(4\pi
|\vr - \vr'|\rt)$ is the Green's function of the Laplacian operator
$\del_{\vr}^2$, we in addition deduce from Eq.~(10d) that,
\de{\del_{\vr}^2\cF_G(\vr) = (2\pi G/c^4)T_{G = 0}(\vr)\cF_G(\vr),}
which is readily reexpressed as a zero-energy stationary-state nonrel%
ativistic \e{Schr\"{o}dinger equation} for the dimensionless wave
function $\cF_G(\vr)$, namely,
\re{
    \lf(-\h^2\del_{\vr}^2/(2m) + V(\vr)\rt)\cF_G(\vr) = 0,
}{11b}
whose repulsive potential $V(\vr)$ is defined as,
\re{
    V(\vr)\eqdf\lf[\pi\h^2G/(mc^4)\rt]T_{G = 0}(\vr).
}{11c}
Because of Eq.~(9c) it is clear from Eq.~(11c) that,
\re{
    {\dy \lim_{|\vr|\rta\infty}}V(\vr) = 0,
}{11d}
which, in turn, implies that the large-$|\vr|$ limit of $\cF_G(\vr)$
that is given by Eq.~(11a) is \e{consistent} with the Eq.~(11b) zero-%
energy \Schr\ equation.

Having established the connection of the Eq.~(10d) \e{gravitational}
integral equation to the Eq.~(11b) \e{zero-energy} stationary-state
\Schr\ equation, we now furthermore note that for stationary states of
\e{positive energy} $E > 0$ this \Schr\ equation becomes,
\re{
    \lf(-\h^2\del_{\vr}^2/(2m) + V(\vr)\rt)\la\vr|\psi_E\ra =
    E\la\vr|\psi_E\ra.
}{12a}
From Eq.~(11d) we see that as $|\vr|\rta\infty$, Eq.~(12a) becomes
simply,
\re{
    \lf(-\h^2\del_{\vr}^2/(2m)\rt)\la\vr|\psi_E\ra = E\la\vr|\psi_E\ra,
}{12b}
whose solutions include \e{all} the dimensionless \e{plane waves} $e^
{i\vp\dt\vr/\h}$ for which $\vp$ satisfies $|\vp|^2 = 2mE$, as well,
of course, as the \e{linear combinations} of these which comprise the
the full set of angularly-modulated outgoing and ingoing free
\e{spherical waves} which have \e{this same scalar wave number} $k =
(2mE)^\hfs/\h$~[9].  Thus we see that Eq.~(12a) has a massive \e{in%
herent solution degeneracy}.  A useful \e{resolution} of this solution
degeneracy can be achieved by \e{reexpressing} Eq.~(12a) in an \e{in%
homogeneous} linear form that \e{can only be satisfied} by the \e{par%
ticular} permissible $|\vr|\rta\infty$ asymptotic behavior which
\e{properly accords} with the \e{design} of a specified experiment.

That idea underlies the Lippmann-Schwinger \e{inhomogeneous modifica%
tion of} Eq.~(12a), which \e{forces} its wave function to behave as a
specifically chosen \e{single} permitted plane wave $e^{i\vp\dt\vr/
\h}$ plus \e{only outgoing} angularly-modulated spherical waves in the
asymptotic region $|\vr|\rta\infty$ where Eq.~(12a) is adequately des%
cribed by Eq.~(12b).  If we denote as $\la\vr|\psi^+_\vp\ra$ the solu%
tion of Eq.~(12a) which \e{satisfies} this \e{particular} permitted
$|\vr|\rta\infty$ asymptotic behavior, then the \e{inhomogeneous}
Lippmann-Schwinger equation that \e{uniquely} describes $\la\vr|\psi^+
_\vp\ra$ is~[10],
\re{
    \la\vr|\psi^+_\vp\ra = e^{i\vp\dt\vr/\h} -
                           \la\vr|(-\h^2\q{\del^2}/(2m)
    -|\vp|^2/(2m) - i\ep)^{-1}\qV|\psi^+_\vp\ra.
}{13a}
From a \e{static gravitational} standpoint the \e{relevant feature} of
the Eq.~(13a) inhomogeneous Lippmann-Schwinger equation and its wave
function $\la\vr|\psi^+_\vp\ra$ is that,
\re{
\lf.\la\vr|\psi^+_\vp\ra\rt|_{\vp = \vz} = \cF_G(\vr),
}{13b}
as is seen from comparison of the $\vp = \vz$ case of Eq.~(13a) with
Eq.~(10d)---to make this comparison one must use Eq.~(11c) to express
$V(\vr)$ as the appropriate constants times $T_{G = 0}(\vr)$, and one
must also use the fact that the integral transform kernel $-1/\lf(4\pi
|\vr - \vr'|\rt)$ is the coordinate-representation inverse (i.e.,
Green's function) of the Hilbert-space ``Laplacian'' operator $\q{\del
^2} = -|\qvp|^2/\h^2$, namely that,
\de{\la\vr|(\q{\del^2})^{-1}|\vr'\ra =
     -1/\lf(4\pi|\vr - \vr'|\rt).}
Note that the \e{negative} imaginary infinitesimal $-i\ep$ which ap%
pears in Eq.~(13a) is \e{unnecessary} when $\vp = \vz$, which repre%
sents a \e{purely static} state of affairs that has \e{no} distin%
guishable outgoing versus ingoing spherical waves.  Indeed the massive
solution degeneracy of the stationary-state \Schr\ equation given by
Eq.~(12a) \e{collapses} when $E = 0$.

Given the Eq.~(13b) close relationship of the local gravitational en%
ergy-reduction factor $\cF_G(\vr)$ to the Lippmann-Schwinger wave
function $\la\vr|\psi^+_\vp\ra$, it would seem logical to apply well-%
known general solution methods for Lippmann-Schwinger equations to our
gravitational Eq.~(10d).  Unfortunately, however, the only widely-ap%
plied fully general solution method for Lippmann-Schwinger equations
is \e{perturbational} in character, and therefore is \e{inherently}
subject to \e{failure}.

\sct{The struggle to transcend the perturbational Born trap}
\noindent
If we bring the second term on the right-hand side of the Eq.~(13a)
Lippmann-Schwinger equation to its left-hand side, we obtain,
\re{
    \la\vr|\psi^+_\vp\ra +
    \la\vr|(\qK - E_\vp - i\ep)^{-1}\qV|\psi^+_\vp\ra =
    e^{i\vp\dt\vr/\h},
}{14a}
where $\qK\eqdf(-\h^2\q{\del^2}/(2m))$ is the kinetic energy
\e{operator} and $E_\vp\eqdf(|\vp|^2/(2m))$ is the kinetic ener%
gy \e{c-number scalar} that corresponds to the c-number momentum vec%
tor $\vp$.  Taking $\la\vr|\vp\ra\eqdf e^{i\vp\dt\vr/\h}$, the formal
solution of Eq.~(14a) is,
\re{
    \la\vr|\psi^+_\vp\ra =
\la\vr|[1 + (\qK - E_\vp - i\ep)^{-1}\qV]^{-1}|\vp\ra.
}{14b}
The only way forward at this point would seem to be expansion of the
inverse of the operator in square brackets in the well-known Born geo%
metric perturbation series, which involves successive \e{powers} with
alternating signs of the particular operator,
\de{\qX\eqdf (\qK - E_\vp - i\ep)^{-1}\qV,}
acting on the momentum eigenstate $|\vp\ra$~[11].  If the operator
$\qX$ dominates the identity on the momentum eigenstate $|\vp\ra$, it
is not unlikely that the Born geometric series \e{diverges}.  One
might suppose that in such instances one could simply \e{recast} the
Born geometric expansion to be in powers of the \e{inverse} of $\qX$,
since one is formally free to choose \e{either} the expansion $[1 +
\qX]^{-1} = 1 - \qX + \qX^2-\cdots$ \e{or} the expansion $[1 + \qX]^
{-1} = \qX^{-1}[1 + \qX^{-1}]^{-1} = \qX^{-1}-(\qX^{-1})^2 + (\qX^{-1}
)^3 -\cdots$.  Most unfortunately, however, since $\qX^{-1} = \qV^{-1}
\lf(\qK - E_\vp\rt)$, the operator $\qX^{-1}$ \e{vanishes altogether
when applied to the momentum eigenstate} $|\vp\ra$.  This unanticipat%
ed abrupt setback is a stark warning that hidden snares beset Born-%
style geometric perturbation expansions for the Lippmann-Schwinger
equation.

For the $\vp = \vz$ case of the Lippmann-Schwinger equation that ap%
plies to \e{static gravitation}, we are, of course, \e{particularly}
interested in \e{arbitrarily large} energy densities $T_{G = 0}(\vr)$,
and therefore in \e{arbitrarily strong} operators $\qV$ and $\qX$.
Thus it is clear that we need to \e{entirely abjure} Born-style geo%
metric perturbation expansion, but how could that \e{conceivably} be
accomplished \e{in practice}?  The \e{only} possibility is through ex%
ploration of \e{nonstandard manipulations} of the Lippmann-Schwinger
equation.

Returning to Eq.~(14a) we now \e{deliberately shun} the \e{elegant and
natural} factorization of the \e{operator} $[1 + \qX]$ on its left-%
hand side, which can only lead us down the primrose path to the Born
geometric perturbation series, and \e{instead} opt to forcibly factor
that side into two \e{mere functions of the vector coordinate} $\vr$,
the \e{first} of which is \e{still} $\la\vr|\psi^+_\vp\ra$ because we
aim for a result which is at least \e{akin} to Eq.~(14b), but the
\e{second} of which is \e{repulsively inelegant}, being \e{merely the 
product of} $(\la\vr|\psi^+_\vp\ra)^{-1}$ with the \e{original
left-hand side of} Eq.~(14a).  There \e{is} in fact ``method'' in that
gross ugliness, however, because we can now \e{actually arithmetically
divide} the plane wave $e^{i\vp\dt\vr/\h}$ on the \e{right-hand side}
of Eq.~(14a) by that gauche \e{second} factor \e{without resorting to
any kind of perturbation expansion}.  A very heavy price has been paid
in the coin of gross inelegance, but the goal of \e{no perturbation
expansion whatsoever} has been achieved.  To be sure, the almost
childish manipulations just described \e{haven't} extracted any final
\e{result} from Eq.~(14a), what they have produced is \e{only} a basis
for \e{refinement through iteration}.  It is readily seen, however,
that the iteration process is \e{devoid of perturbational characteris%
tics}; it \e{instead} resembles a \e{continued fraction}.

The iteration formula we have just extracted from Eq.~(14a) is expli%
citly,
\re{
    \la\vr|\psi^{(n + 1)+}_\vp\ra = e^{i\vp\dt\vr/\h}/[1 +
    (\la\vr|\psi^{(n)+}_\vp\ra)^{-1}
\la\vr|(\qK - E_\vp - i\ep)^{-1}
     \qV|\psi^{(n)+}_\vp\ra],
}{14c}
for $n = 0, 1, 2,\ldots$, where, of course, $\la\vr|\psi^{(0)+}_\vp\ra
= e^{i\vp\dt\vr/\h}$.  Therefore $(\la\vr|\psi^{(0)+}_\vp\ra)^{-1}$ is
obviously well-defined, and the form of Eq.~(14c) makes it apparent
that for $n = 1, 2,\ldots$, $(\la\vr|\psi^{(n)+}_\vp\ra)^{-1}$ is
well-defined as well.  That the iteration formula of Eq.~(14c) does
not have perturbational characteristics, but rather those of a contin%
ued fraction is also manifest.

Finally, on inserting $\vp = \vz$ into Eq.~(14c) we obtain the itera%
tion formula for the gravitational energy-reduction factor $\cF_G(\vr
)$, which is,
\re{
    \cF^{(n + 1)}_G(\vr) = 1/[1 + \hf(G/c^4)(\cF^{(n)}_G(\vr))^{-1}
    \int|\vr - \vr'|^{-1}T_{G = 0}(\vr')\cF^{(n)}_G(\vr')d^3\vr'],
}{15a}
for $n = 0, 1, 2,\ldots$, where, of course, $\cF^{(0)}_G(\vr) = 1$.

Since $T_{G = 0}(\vr)\ge 0$ from Eq.~(9a), $T_{G = 0}(\vr)$ is smooth
from Eq.~(9b), and $\int T_{G = 0}(\vr')d^3\vr' < \infty$ from Eq.~%
(9c), it is clear from Eq.~(15a) that,
\re{
  \hbox{if $1\ge\cF^{(n)}_G(\vr) > 0$, then $1\ge\cF^{(n + 1)}_G(\vr) > 0$.}
}{15b}
Therefore we can conclude that,
\re{
    1\ge\cF_G(\vr) > 0.
}{15c}
From Eqs.~(9a), (10b) and (15c) we can deduce that,
\re{
    0\le T_G(\vr)\le T_{G = 0}(\vr).
}{15d}
From Eq.~(10c) and the fact that the gravitational energy-reduction
factor $\cF_G(\vr)$ satisfies $\cF_G(\vr) > 0$ we can deduce that,
\de{{\dy 2(c^4/G) > \int|\vr - \vr'|^{-1}T_G(\vr')d^3\vr' = 
\int|\vr''|^{-1}T_G(\vr + \vr'')d^3\vr'',}}
where $\vr''\eqdf(\vr' - \vr)$.  Furthermore, we have that,
\de{{\dy \int|\vr''|^{-1}T_G(\vr + \vr'')d^3\vr'' \ge
    \int_{|\vr''|\le R}|\vr''|^{-1}T_G(\vr + \vr'')d^3\vr'' \ge
    R^{-1}\int_{|\vr''|\le R}T_G(\vr + \vr'')d^3\vr''.}}
Therefore from the two foregoing lines of displayed integral inequali%
ties we can conclude that,
\re{
    {\dy (c^4/G)(2R) > \int_{|\vr''|\le R}T_G(\vr + \vr'')d^3\vr'',}
}{15e}
namely that the self-gravitationally corrected static energy contained
in any sphere cannot exceed the diameter of that sphere times
$(c^4/G)$.  Note that this result is \e{completely independent of any
assumption concerning symmetry properties of the energy distribution},
and that it is in agreement with the extremely nonspherical result for
two discrete particles which is given by Eq.~(1b).  Therefore it is
likely to be overly conservative in practice.  A more practical energy
upper-bound estimate can be obtained by simply averaging the maximum
possible and minimum possible values of $|\vr''|$ that occur when
integrating over the sphere of radius $R$, which yields $R/2$, and
therefore the ``rough'' bound,
\de{
    {\dy (c^4/G)R\; \widetilde{>} 
     \int_{|\vr''|\le R}T_G(\vr + \vr'')d^3\vr'',}}
which is in line with the result of Eq.~(4b), for which spherical sym%
metry was assumed.

No doubt the most interesting aspect of what has just been presented
here is the unfastening of the shackles of the perturbational Born-ex%
pansion paradigm for a class of equation systems that incorporate lin%
ear operators.  A superior iteration method has been developed by del%
iberately shunning an attractive natural relationship that involves
those linear operators in favor of concocting a clumsy artificial re%
lationship that involves \e{only function values}.  The point of pro%
ceeding in this way is that purely \e{arithmetic} operations with
\e{function values} require \e{no approximations}, whereas even quite
elementary-looking operations involving \e{operators} may \e{not} be
practically feasible without making use of potentially disastrous per%
turbation expansions.  Indeed function manipulations can, on the con%
trary, be \e{directed} toward the goal of achieving iteration schemes
that have \e{continued fraction} rather than perturbational character.

\sct{Conclusion}
\noindent
We have constructed a spherically-symmetric Newtonian gravitostatic
model with energy feedback that yields a simple and apparently very
useful upper bound on the amount of self-gravitationally corrected
energy which can be contained in a spherical region.  That bound is
$(c^4/G)$ times the radius $r_s$ of the sphere, a relationship which,
inter alia, implies that the Schwarzschild radius \e{never} lies in
free space, making the Schwarzschild singularity physically
unrealizable. We have, moreover, been able to extend this Newtownian
gravitostatic model to \e{nonsymmetric} energy distributions by means
of an interesting nonperturbational continued-fraction iteration-%
solution method.  That extension turns out to yield \e{qualitatively
the same upper bound} (i.e., to within a factor of two) on the amount
of self-gravitationally corrected static energy which can be contained
in a spherical region.

Such a $(c^4/G)$ times radius upper bound on a sphere's contained
energy cuts off the mass deviation of an interacting quantum virtual
particle at approximately the Planck mass, which in principle does
away with the ultraviolet divergences that bedevil quantum particle
scattering amplitude calculations.

The $(c^4/G)$ times radius bound on a sphere's contained energy
ought to be \e{attained} for contained quantum fields, which have
\e{infinite energy} before self-gravitational correction, due to the
combination of their unbounded frequency spectra and the quantum un%
certainty principle.  But \e{only} the universe \e{itself}, with its
cosmological redshift, is actually capable of ``containing'' the arbi%
trarily high frequencies of a quantum field.  Roughly estimating the
radius of the universe as its age times the speed of light, and then
dividing the $(c^4/G)$ times this radius by the corresponding
spherical volume that has this radius, yields a rough averaged ``dark
energy'' density estimate of about 1.7 joules per cubic kilometer,
which is of the same order of magnitude as observational data.  The
\e{same} formula suggests that the \e{early} universe might have had
an immensely greater ``dark energy'' density, perhaps as much as a
Planck unit of energy density, which would be roughly 120 orders of
magnitude times its present value.  It is interesting that this seems
to provide an \e{automatic} mechanism for the inflation of the early
universe.  The formula also suggests that the ``dark energy'' density
will be decreasing toward zero as the universe expands, but that it
won't decrease as rapidly as the density of ordinary matter will,
which will increase the relative dominance of dark energy.

Finally, there seems to be \e{no reason} why the \e{presently evolved}
``dark energy'' density \e{should not share the small inhomogeneities
which are so typical of the rest of the universe}, such as the small
peaks in the cosmic microwave background, and the galaxies, groups,
filaments and voids in the distribution of luminous matter.  If ``dark
energy'' indeed has inhomogeneities, then might not those inhomogenei%
ties \e{themselves} be the thing we call ``dark matter''?  In spite of
all the gravitational evidence for ``dark matter'', there is apparent%
ly no observationally-known non-gravitational signal whatsoever for
it.  It would be a relief if something so elusive ultimately turned
out to \e{not} have an independent existence.

\vskip 1.75\baselineskip\noindent{\frtbf References}
\vskip 0.25\baselineskip

{\parindent = 15pt
\sk\item{[1]}
L. I. Schiff,
\e{Quantum Mechanics}
(McGraw-Hill, New York, 1955), p.~140, Eq.~(23.29).
\sk\item{[2]}
C. Schiller,
International Journal of Theoretical Physics {\bf 44},
1629--1647 (2005);
arXiv:physics/0607090 [physics.gen-ph] (2006).
\sk\item{[3]}
S. Weinberg,
\e{Gravitation and Cosmology: Principles and Applications of the
General Theory of Relativity}
(John Wiley \& Sons, New York, 1972), p.~80, Eq.~(3.5.3).
\sk\item{[4]}
S. Weinberg, op.\ cit., p.~181, Eq.~(8.2.14).
\sk\item{[5]}
S. Weinberg, op.\ cit., Section~3.6, pp.~85--86.
\sk\item{[6]}
Wikipedia,
``Age of the universe'',
http://en.wikipedia.org/wiki/Age\_of\_the\_universe.
\sk\item{[7]}
Wikipedia,
``Dark energy'',
http://en.wikipedia.org/wiki/Dark\_energy.
\sk\item{[8]}
S. M. Carroll,
arXiv:astro-ph/0310342 (2003);
AIP Conf.\ Proc.\ {\bf 743},
16--32 (2005).
\sk\item{[9]}
L. I. Schiff, op.\ cit., Sec.~15, pp.~77--79.
\sk\item{[10]}
M. L. Goldberger and K. M. Watson,
\e{Collision Theory}
(John Wiley \& Sons, New York, 1964), pp.~197--199.
\sk\item{[11]}
M. L. Goldberger and K. M. Watson, op.\ cit., pp.~306--313.
}
\bye